\documentclass[aps,prl,twocolumn,superscriptaddress]{revtex4-1}%
\usepackage{graphicx}

\usepackage{units}
\usepackage{amsmath}
\usepackage{textcomp}
\newcommand{\cl}{$^{35}$Cl\,}

\begin{document}


\title{ARPES sensitivity to short-range antiferromagnetic correlations.} 



\author{R. Wallauer}
\affiliation{Technion - Israel Institute of Technology, Department of Physics, Technion City, 32000 Haifa, Israel}

\author{S. Sanna}
\affiliation{Dipartimento di Fisica and Unit� CNISM di Pavia, I-27100 Pavia, Italy}

\author{E. Lahoud}
\affiliation{Technion - Israel Institute of Technology, Department of Physics, Technion City, 32000 Haifa, Israel}

\author{P. Carretta}
\affiliation{Dipartimento di Fisica and Unit� CNISM di Pavia, I-27100 Pavia, Italy}

\author{A.Kanigel}
\affiliation{Technion - Israel Institute of Technology, Department of Physics, Technion City, 32000 Haifa, Israel}


\date{\today}

\begin{abstract}Angle-resolved photoemission spectroscopy (ARPES) is one of most
powerful techniques to unravel the electronic properties of
layered materials and in the last decades it has lead to a
significant progress in the understanding of the band structures
of cuprates, pnictides and other materials of current interest. On
the other hand, its application to Mott-Hubbard insulating
materials where a Fermi surface is absent has been more limited.
Here we show that in these latter materials, where electron spins
are localized, ARPES may provide significant information on the
spin correlations which can be complementary to the one derived
from neutron scattering experiments.
Sr$_2$Cu$_{1-x}$Zn$_x$O$_2$Cl$_2$, a prototype of diluted spin
$S=1/2$ antiferromagnet (AF) on a square lattice, was chosen as a
test case  and a direct correspondence between the amplitude of
the spectral weight beyond the AF zone boundary derived from ARPES
and the spin correlation length $\xi$ estimated from $^{35}$Cl NMR
established. It was found even for correlation lengths of a few
lattice constants a significant spectral weight in the back-bended
band is present which depends markedly on $\xi$. Moreover the
temperature dependence of that spectral weight is found to scale
with the $x$ dependent spin-stiffness. These findings prove that
ARPES technique is very sensitive to short-range correlations and
its relevance in the understanding of the electronic correlations
in cuprates is discussed.
\end{abstract}


\maketitle 


The discovery of high temperature superconductivity in the
cuprates has promoted a renewed interest for the study of electron
spin correlations in quantum $S=1/2$ spin systems. Inelastic
neutron scattering (INS) \cite{Keimer1992} and nuclear spin-lattice relaxation rate ($1/T_1$)
\cite{Rigamonti1998} measurements have been widely employed to investigate the
evolution of spin correlation in several layered materials as
cuprates, manganites and pnictides \cite{Tucker2012} and a series of major breakthroughs
obtained. Nevertheless, each one of these techniques has its
drawbacks. Inelastic neutrons scattering requires large crystals
which are not always available while the determination of the spin
correlation length $\xi$ from nuclear magnetic resonance (NMR)
$1/T_1$ is not direct but rather based on scaling assumptions.
Only when scaling exponents are known, as it is the case for the
cuprates, also NMR $1/T_1$ allows a quantitative estimate of
$\xi$.\cite{Carretta1997}

Remarkably, the onset of sizeable antiferromagnetic (AF)
correlations can affect also the band structure. In fact, due to
the AF order the Brillouin zone boundary is reduced to half of its
size in the ordered state and electron bands bend-back at the AF
zone boundary. Hence, it is tempting to use a technique which is
quite sensitive to the details of the band features, as it is the
case of angle-resolved photoemission spectroscopy (ARPES), to
investigate the electron spin correlations.  ARPES has allowed to
clarify how the Fermi surface of high $T_c$ superconductors (HTSC)
evolves with hole doping and which parts are likely involved in
the pairing mechanisms, \cite{Damascelli2003} becoming one of the
most attractive techniques to study electronic correlations in
layered materials. On the other hand, the Charge Density Wave
(CDW) instabilities recently detected in HTSC
\cite{CDW3,CDW2,CDW4} are somewhat elusive to ARPES since often
they do not give rise to the typical "finger-prints" of CDW
order\cite{Veronic,Elias}. For example, the change in the Fermi
momentum observed in Bi2201 \cite{Hashimoto} which could result
from a CDW phase is not observed in Bi2212 \cite{PRL08}.
Accordingly it has been argued that the absence of a signature of
CDW order in ARPES spectra could be likely due to the short range
or even dynamic character of the CDWs. Here we show that, on the
contrary, ARPES is sensitive to short range electron spin
correlations and that it may represent a sensitive probe,
complementary to INS, particularly when the correlation length
approaches a few lattice steps.

We have chosen Sr$_2$CuO$_2$Cl$_2$ (SCOC), one of the best
experimental realization of a spin $S=1/2$ two-dimensional (2D)
antiferromagnet on a square lattice, as a test case for ARPES
sensitivity to short range spin correlations. This material is
characterized by a well defined order parameter and magnetic
wave-vector $\vec{q}$. Moreover, its in-plane AF correlation
length, $\xi$, above the N\'eel temperature $T_N$ was studied using INS \cite{Greven1994} and $^{35}$Cl NMR \cite{Borsa1992}. In
addition due to the presence of a natural cleaving plane SCOC is
perfectly suited for ARPES experiments.

\begin{figure}
    \includegraphics[scale=1]{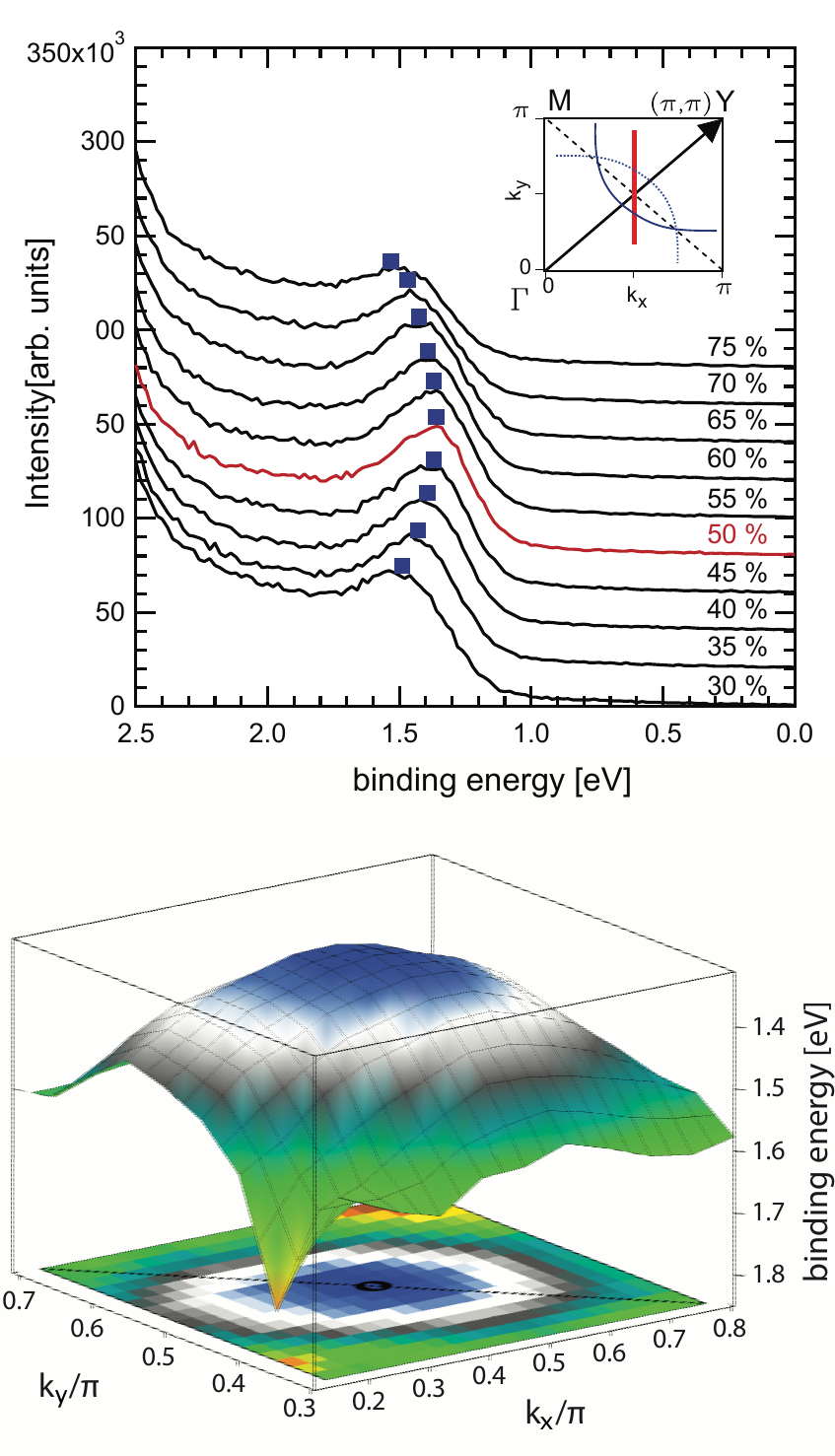}
    \caption{\label{fig:Dispersion_withInset} Upper panel: Band dispersion along the  $\Gamma$ - M direction around the $(\pi/2,\pi/2)$-point. The position of the peak in each EDC is indicated by blue dots and we find the lowest binding energy at the AFM zone boundary (red line). Inset: A typical cuprate Fermi surface and the folded band upon antiferromagnetic ordering with wave-vector $\vec{q} = (\pi,\pi) $ (black arrow). Lower panel: Extracted dispersion from peak position of several scans in the $\Gamma$ - M direction. The top of the band moves from the zone corner to ($\pi/2,\pi/2$) as indicated by the black circle in the 2D projection.}%
\end{figure}

Many ARPES studies have been carried out on SCOC \cite{Wells1995,
LaRosa1997, Pothuizen1997, Durr2000, Haffner2000, Haffner2001} and
on the nearly identical Ca$_2$CuO$_2$Cl$_2$ system
\cite{Ronning1998, Ronning2005}, both below and above T$_N$. As
expected,  owing to the AF ordering the Brillouin zone boundary is
reduced to half of its size in the ordered state and the band
bends-back at the antiferromagnetic zone boundary (see inset in
Fig. \ref{fig:Dispersion_withInset}). Remarkably, even if this
back-bending is believed to be a result of the AF order it
persists up to temperatures well above T$_N $= 256 K. This can be
explained by the very long AF correlations which are about 100
lattice constants at 300 K \cite{Greven1994}. In this work we were
able to reduce $\xi$ down to a few lattice constants $a$, measure
with ARPES the intensity in the back-bended part and relate it to
a change in the correlation length estimated from $^{35}$Cl NMR
$1/T_1$ measurements. As it will be shown in the following some
intensity in the back-bended part persists even for $\xi/a\simeq
3$. This is the first systematic study of the sensitivity of ARPES
to the AF correlation length which can be relevant also for the
understanding of other fluctuating orders.

SCOC single crystals were grown from the melt according to the
procedure described by Miller et al. \cite{Miller1990}. All ARPES
measurements were done with 21.2 eV photons  from a monochromized
He lamp and a Scienta R4000 electron analyser. Due to the broad
band under investigation and for the sake of better statistics,
spectra were typically recorded with 20 meV energy resolution and
0.2 degree angular resolution. Fig. \ref{fig:Dispersion_withInset}
shows the dispersion of the highest occupied band at 300 K in the
direction indicated by the red line in the inset. A minimal
binding energy of about 1.4 eV is found exactly at the AF zone
boundary, indicated by the red curve. Beyond that point the band
bends-back toward higher binding energies, as expected in view of
the new periodicity of the system. At high binding energies around
the zone centre  we find a waterfall-like feature, typical of many
cuprates, whose origin is still under debate
\cite{Inosov2007,Basak2009,Graf2007}. We note that since  SCOC is
insulating, the chemical potential is set by impurity levels and
can vary significantly between different samples. For that reason
the gap size cannot be measured.

In agreement with previous works we find no sharp peaks even at
the lowest binding energy. The origin of the line broadening still
needs to be clarified since numerical calculations based on the
t-J-model actually predict sharp quasiparticle-like
excitations.\cite{Pothuizen1997,
Kim2002,Laughlin1997,Shen2004,Sangiovanni2006,Mishchenko2004} By
measuring several momentum cuts and extracting the peak positions
for each Energy Distribution Curve (EDC), we were able to map the
dispersion of the band over a large area of the Brillouin zone (see Fig \ref{fig:Dispersion_withInset} lower panel).
The dispersion is symmetric around the new zone boundary and has a
maximum at ($\pi/2,\pi/2$). We repeated the measurements at
various temperatures between 77 K, deep in the AF ordered phase
and up to 450K, nearly twice T$_N$. We find almost no change in
the spectra and, in particular, at all temperatures we observe the
back-bended band beyond ($\pi/2,\pi/2$). However, it should be
noticed that in SCOC even at 450K $k_BT\ll J$, the in-plane
superexchange coupling, and the correlation length is still more
than 20 lattice constants \cite{Greven1994}.

In order to reduce the correlation length we partially substituted
$S=1/2$-Cu$^{2+}$-ions by spinless Zn$^{2+}$-ions. In fact, it is
well established that in La$_2$Cu$_{1-x}$(Zn,Mg)$_x$O$_4$ T$_N$
can be reduced by diluting the spin lattice
\cite{Carretta1997,Vajk2002,Carretta2011} and long-range order can
even be completely suppressed for substitution levels beyond the
percolation threshold (41\% for a square lattice). More important
for our needs, Zn can significantly reduce the AF correlation
length $\xi$ above T$_N$. We found that Zn can substitute Cu in
SCOC crystals up to the substitution limit of 29\%, a value below
the percolation threshold but large enough to yield a significant
reduction of the AF correlations and of T$_N$. In particular, at
this substitution level T$_N$ is reduced to 180 K.

\begin{figure}
\includegraphics[scale=1]{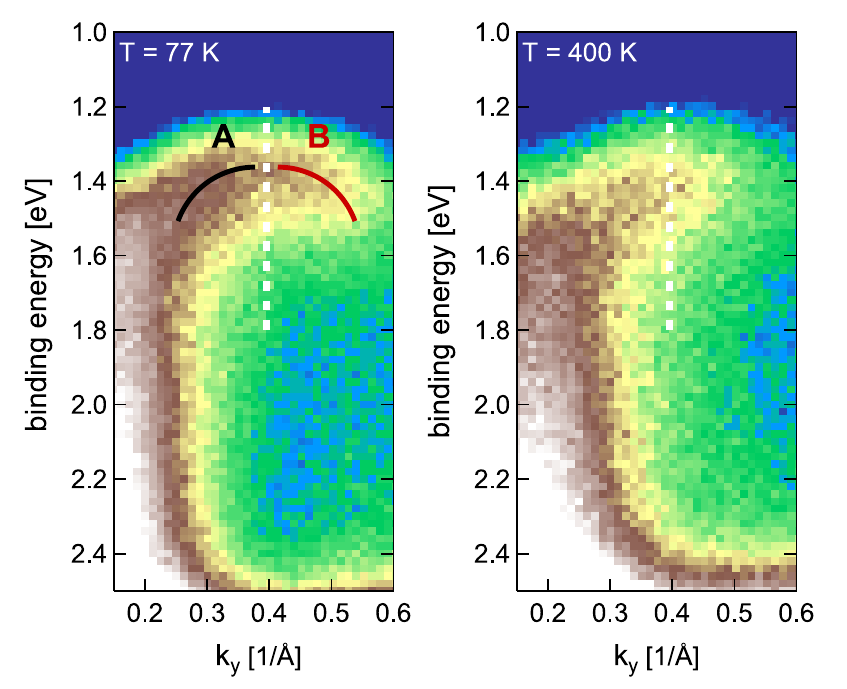}
\caption{\label{fig:compareBand}
Comparison of the ARPES images for the  x=0.29 substituted sample at 77K (left) and at 400K (right). The loss of spectral weight beyond the ($\pi/2,\pi/2$)-point, indicated by the white line, can be observed at high temperature. Nevertheless, the back-bended band is clearly visible, although the correlation length at this temperature is 2-3 lattice constants. The black (A) and red (B) lines represent the band dispersion on both sides of the new zone-boundary at ($\pi/2,\pi/2)$  }%
\end{figure}

The ARPES spectra of the the spin diluted samples, at all
temperatures and at all Zn concentrations, still show intensity
beyond the AF zone boundary. In particular, in Fig.
\ref{fig:compareBand} we show two normalized ARPES images for the
x=0.29 sample, each spectrum is normalized using the intensity at
high-binding energy around the $\Gamma$ point and a constant
background, measured at low binding-energy in the Mott gap, was
subtracted. In the left panel of Fig. \ref{fig:compareBand} we
show an image measured at T=77K in the AF state and on the right
the corresponding image at T=400 K, where the correlation length
is only 3 lattice constants based on the high temperature
extrapolation of the NMR data. One can clearly see that, even if
at 400K the intensity of the back-bended part is much weaker than
at 77 K, there is still some intensity beyond the AF zone
boundary.

In order to provide a quantitative estimate of that intensity we
performed the following analysis.  We integrate the intensity
along the dispersion maximum on both sides of the
$(\pi/2,\pi/2)$-point (dashed white line in Fig.
\ref{fig:compareBand}) as indicated by the black (A) and red (B)
lines. The range of the integration area is 300 meV along the
energy axis. Along the momentum axis we choose a range that starts
at a point near the $\Gamma$-point, where the dispersion is not
affected by the waterfall-like feature, and ends at the
$(\pi/2,\pi/2)$-point. An identical range was chosen on the other
side of the $(\pi/2,\pi/2)$-point. Thus, the ratio of the
integrals on both sides, B/A, provides a measure of the loss of
spectral weight beyond the AF zone boundary.

\begin{figure}
    \includegraphics[scale=1]{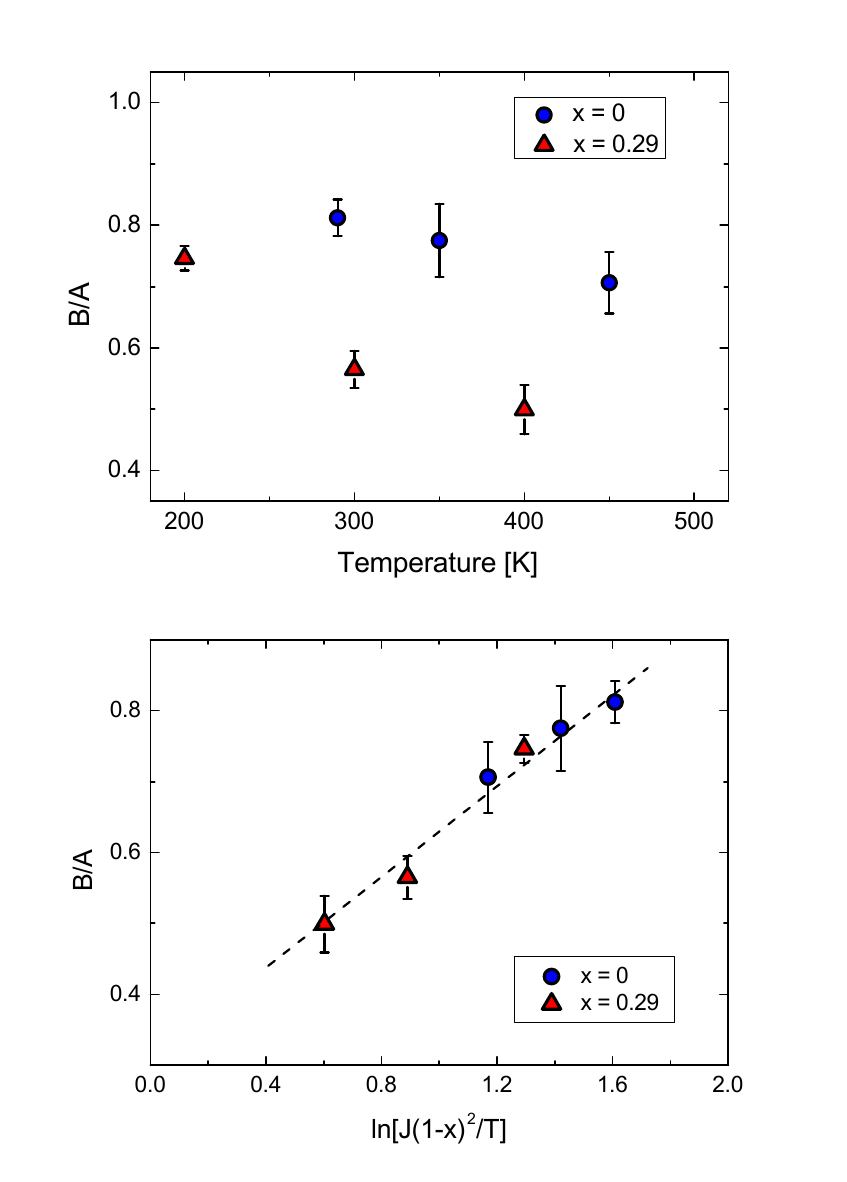}
    \caption{\label{fig:corrlenABRatio} (a) The ratio of the integrated intensity along the black (A) and red (B)
    lines in figure \ref{fig:compareBand} as function of the temperature. (b) The ratio B/A is reported as a function
    of $ln[J(1-x)^2/T]$ in order to evidence the scaling of its $T$-dependence with the spin-stiffness.}%
\end{figure}

In Fig. \ref{fig:corrlenABRatio}(a) we show the ratio B/A as a
function of temperature for the pristine and for the  x=0.29
samples. For the pristine sample the ratio B/A decreases rather
slowly with temperature, which is not surprising if one considers
that this 2D AF is characterized by a $J\simeq 1450$ K. On the
other hand, in the Zn substituted sample this ratio decreases
faster with temperature. In a spin diluted $S=1/2$ Heisenberg AF
on a square-lattice the T-dependence of the in-plane correlation
length $\xi\propto exp(2\pi\rho_s(x)/T)$ is determined by the
spin-stiffness $\rho_s(x)= 1.15 J(1-x)^2/2\pi$
\cite{Carretta1997,Vajk2002}, where the factor $(1-x)^2$ accounts
for the probability to find two neighbouring electron spins
coupled via the exchange coupling $J$. Accordingly, one should
expect that if B/A ratio is a measure of the spin correlations it
should scale as $J(1-x)^2/T$. Remarkably in Fig.
\ref{fig:corrlenABRatio}(b) one observes that such a scaling does
occur and that B/A$\propto ln[J(1-x)^2/T]$. The understanding of
this logarithmic functional form goes beyond the aim of the
present work.

In order to further evidence the direct relationship between B/A
ratio and the in-plane correlation length we quantitatively
estimated $\xi$ for different spin diluted samples. $\xi$ can be
conveniently derived from nuclear spin-lattice relaxation rate
$1/T_1$, an approach successfully used in the analogous
La$_2$Cu$_{1-x}$Zn$_x$O$_4$ compound \cite{Carretta1997} where a
quantitative agreement with the $\xi$ directly measured by INS was
found. In fact, by resorting to scaling arguments, once the
hyperfine coupling $A$ and dynamical scaling exponent $z$ are
known one can establish a one-to-one correspondence between
$1/T_1$ and $\xi$.\cite{Carretta1997} In particular, if one
considers a 2D $S=1/2$ antiferromagnet on a square lattice and a
nucleus which is essentially coupled just with its nearest
neighbour $S=1/2$ spin, one can write
\begin{equation}
\frac{1}{T_1}= \frac{\gamma^2 A^2}{4} \frac{\hbar\sqrt{\pi}}{Jk_B}
{\xi^{z+2}}\beta(\xi) \int_{BZ} \frac{d\vec q}{(1+q^2\xi^2)^2}  ,
\label{Eq:sec}
\end{equation}
with $\gamma$ the nuclear gyromagnetic ratio and $1/\beta(\xi)=
\xi^2\int_{BZ} {d\vec q}/(1+q^2\xi^2)$ a factor which preserves
the spin sum rule.\cite{Carretta1997} For $\xi\gg 1$ this
expression yields to $1/T_1\propto \xi^z$. In SCOC one can assume
the dynamical scaling exponent $z=1$ derived from INS
\cite{Greven1994} and NQR experiments \cite{Carretta1997} for
other cuprates on a square lattice, as La$_2$CuO$_4$. Since $A$
was not very precisely estimated from previous shift
measurements,\cite{Borsa1992} we decided to determine it by taking
for $A$ the value which allows to match the absolute value for
$\xi$ determined from \cl $1/T_1$ and from INS over the same T
range. Such a match occurs for $A\simeq 2.7$ kOe, a value very
close to the one that can be derived from the high-T limit of
$1/T_1$. In the case of diluted samples one has to consider a
correction of the average hyperfine coupling by a factor $(1-x)$
\cite{Carretta1997}, accounting for the probability that \cl
nuclear spin is coupled to the nearest neighbour Cu$^{2+}$ spin.

\begin{figure}[h!]
\includegraphics[scale=1]{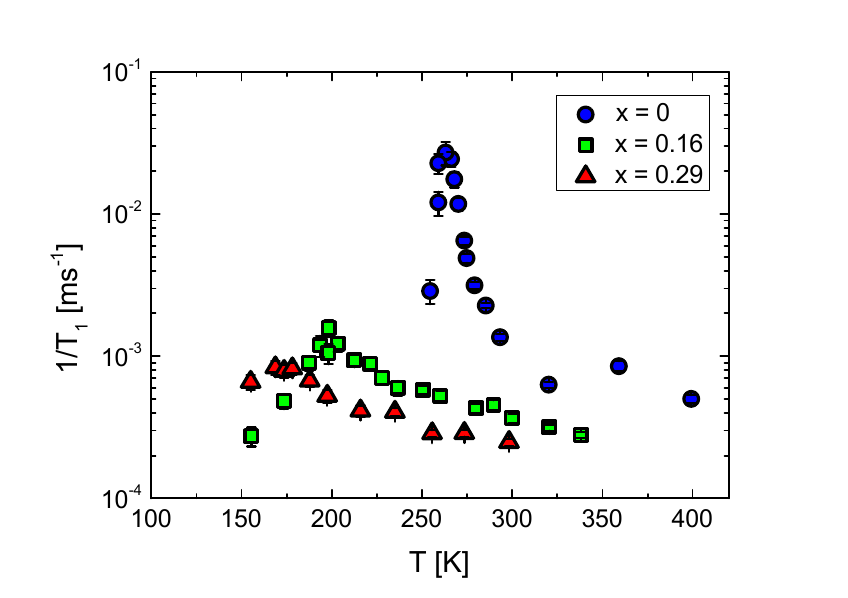}
\caption{Temperature dependence of \cl  $1/T_1$ in
Sr$_2$Cu$_{1-x}$Zn$_x$O$_2$Cl$_2$ for $x=0, 0.16, 0.29$ measured
in a 5.6 Tesla magnetic field. T$_N$ is reduced by the Zn
substitution as can be measured by following the peak position in
$1/T_1$ as the Zn concentration is increased. }\label{Fig:T1vsT}
\end{figure}

\begin{figure}
\includegraphics[scale=1]{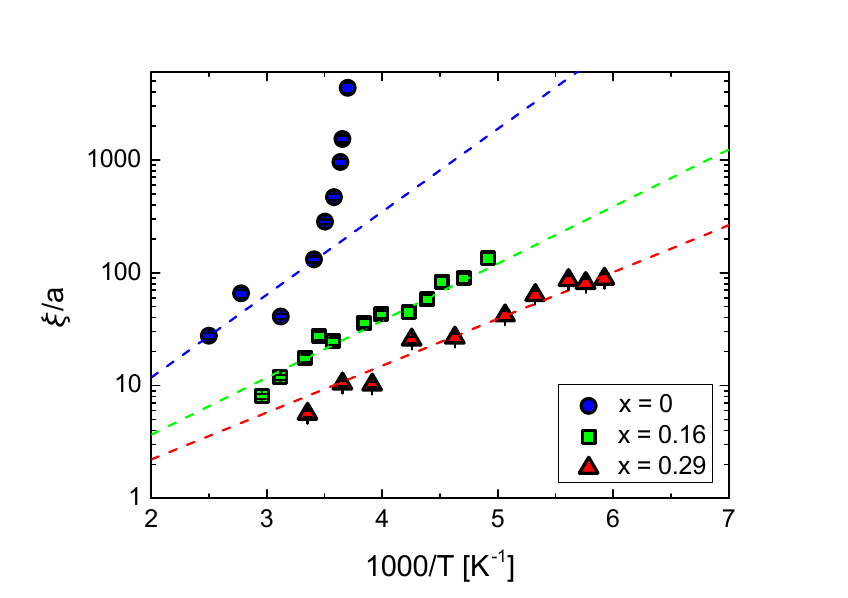}
\caption{\label{fig:corrlen_NMR} The AF correlation length in
units of lattice constants as function of $1/T$ for three
different Zn concentrations. $\xi$ was calculated from the $1/T_1$
data using Eq. \ref{Eq:sec}. The dashed lines show the exponential
growth expected for a two-dimensional  $S=1/2$ antiferromagnet.
The deviation from that trend observed for $x=0$ at $T\rightarrow
T_N$ was ascribed to a small planar spin
anisotropy \cite{Borsa1992}. }
\end{figure}

The temperature dependence of \cl $1/T_1$, determined by using
standard radiofrequency pulse sequences, is shown in Fig.
\ref{Fig:T1vsT} for three representative samples. One observes
that $1/T_1$ progressively increases upon cooling, shows a peak at
T$_N$  and it decreases upon further cooling down. The progressive
growth of $1/T_1$ upon approaching T$_N$ from above is associated
with the progressive growth of the in-plane AF correlations (see
Eq.\ref{Eq:sec}).\cite{Borsa1992} One can now use  Eq.
\ref{Eq:sec} to derive $\xi(T,x)$ from $1/T_1$ data. The
temperature dependence of $\xi(x,T)$ is shown in Fig.
\ref{fig:corrlen_NMR}. One notices that for the diluted samples
the correlation length grows exponential on decreasing
temperature, while for the undoped SCOC a crossover to an XY
behaviour, characterized by a more rapid increase of $\xi$, is
observed on approaching T$_N$ as pointed out by previous NMR
experiments.\cite{Borsa1992}

\begin{figure}
    \includegraphics[scale=1]{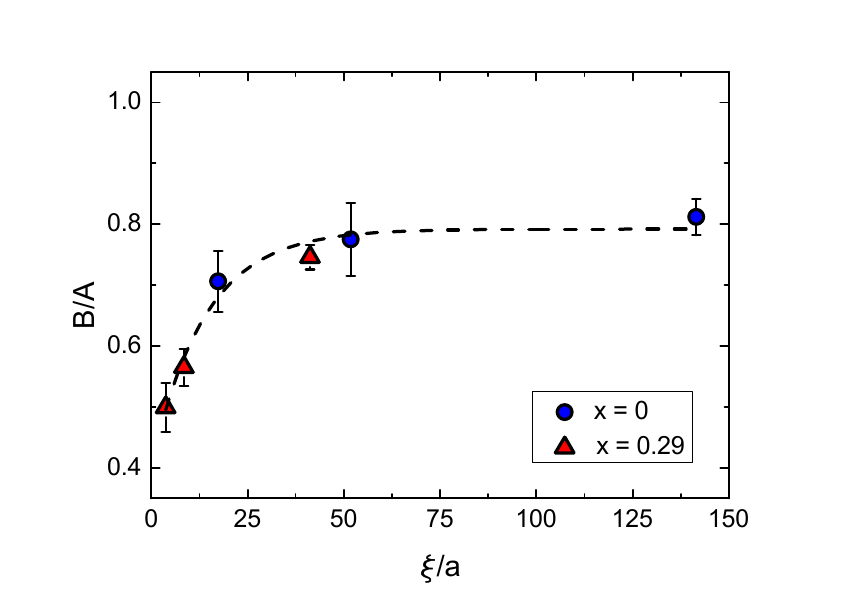}
    \caption{\label{fig:corrT1ABRatio}  The ratio B/A is reported as a function of correlation
    length as measured by NMR data (see figure \ref{fig:corrlen_NMR}). Dashed line is a guide to the eye.}%
\end{figure}

In Fig. \ref{fig:corrT1ABRatio} we show the dependence of ARPES
intensity B/A ratio on the correlation length determined from \cl
$1/T_1$. We find that for both samples the intensity beyond the AF
zone boundary as measured by ARPES is controlled by the AF
correlation length. For $\xi\leq 15$ lattice units the B/A ratio
shows a significant dependence on $\xi$, which clearly
demonstrates that ARPES spectra are indeed sensitive to AF
correlations even when they are as short as a few lattice
constants. This
is an evidence that ARPES besides being a well established
technique to investigate  the electronic structure of layered
materials it allows  also to detect short-range electron spin
correlations. For correlation lengths larger than 15 lattice
constants we observe almost no change in the spectra and hence it
is not possible to distinguish the AF long range order from a
fluctuating phase characterized by $\xi$ of the order of tens of
lattice constants.

In conclusion, we have shown for the first time that the back-bended part of the highest occupied band changes upon decreasing correlation length. This can be seen as a proof that  AF correlations are indeed responsible for this back-bending and that ARPES is a sensitive probe to very short-range AF correlations. Using Zn substitution and by suitably increasing the temperature we were able to control the correlation length down to 3 constants verified by NMR measurements. Even for the shortest correlation length the spectra surprisingly showed significant spectral weight beyond the AF zone boundary. 

The ARPES ability to detect short-range correlations has relevant
implications also for other kind of orders. In particular, our
findings suggest that the absence of the expected fingerprints of
a CDW instability in the ARPES spectra of the under doped cuprates
are not simply a consequence of the short range of the CDW
correlations.

R.W. acknowledges support by the BMBF through the Minerva Fellowship program.

\bibliography{SCOC_bib}

\end{document}